\documentclass[aps,preprint,noshowpacs,nofootinbib,superscriptaddress,groupedaddress]{revtex4}  
\usepackage{graphicx} 
\usepackage{dcolumn}   
\usepackage{bm}        
\usepackage{amssymb}

\begin{document}

\begin{flushright}
CERN-TH-2019-021\\
HIP-2019-03/TH\\
INT-PUB-19-007\\
\end{flushright}

\title{Gluon Radiation from a Classical Point Particle}

\author{K. Kajantie}
\affiliation{Helsinki Institute of Physics, FI-00014 University of Helsinki, Finland}
\email{keijo.kajantie@helsinki.fi}
\author{Larry D. McLerran}
\affiliation{Institute for Nuclear Theory, University of Washington, Box 351550, Seattle, WA, 98195, USA}
\email{mclerran@me.com}
\author{Risto Paatelainen}
\affiliation{Theoretical Physics Department, CERN, CH-1211 Gen\`eve 23, Switzerland}
\email{risto.sakari.paatelainen@cern.ch}
\affiliation{Helsinki Institute of Physics, FI-00014 University of Helsinki, Finland}

\date{\today}

\begin{abstract}
We consider an initially at rest colored particle which is struck by an ultra-relativistic nucleus. The particle is treated classically 
both with respect to its motion and its color charge. The nucleus is treated as a sheet of colored glass within the 
context of the Color Glass Condensate framework. We compute both the momentum and coordinates of the struck
classical particle and the emitted radiation.  Our computations generalize the classic electrodynamics computation of the radiation of an accelerated charged particle to include the radiation induced by the charged gluon field.  This latter contribution adds to the classic electrodynamics result and  produces a gluon rapidity distribution that is roughly constant as a function of rapidity at rapidities far from the fragmentation region of the struck particles. These computations may form the basis of a first principles treatment for the initial conditions  for the evolution of matter produced
in the fragmentation region of asymptotically high energy collisions. 
\end{abstract}


\maketitle

\section{\label{sec:intro}Introduction}

This article will study gluon production in the target fragmentation region of a very high energy hadronic
collision. The goal is to reformulate the very early phenomenological ideas of Anishetty, Koehler and McLerran \cite{Anishetty:1980zp},
see also \cite{km,mclerran,McLerran:2018axu}, in the spirit of the theory of Color Glass Condensate (CGC) \cite{mv}.  
These descriptions make use of the effects of gluon saturation to provide a computational framework for the treatment of high energy QCD processes \cite{Mueller:1985wy, Gribov:1984tu, Mueller:1989st}.

In Ref. \cite{Anishetty:1980zp}, one considered a situation in which an ultrarelativistic nucleus collides with a hadron of radius $R_A$ at rest and imparts a
rapidity $y$ (longitudinal velocity = $\tanh y$) to each of the quarks in the stationary target. A simple calculation then shows that while
the whole system has been accelerated to rapidity $y$, its rest frame length is $e^{-y}\,\,2R_A$, it has been compressed by a factor $e^y$. In Ref. \cite{km},
this estimate was converted into initial conditions of hydrodynamic evolution of energy-momentum and baryon number.  Very recently,
in  Ref. \cite{McLerran:2018avb} it was shown that this compression argument holds quantitatively also quantum mechanically in the
CGC picture.

In addition to compressing the baryonic system, the acceleration of the system will cause radiation of gluons. As an initial stage of
computing this process we shall in this article consider the problem of a single color charged particle interacting with a sheet of 
Color Glass Condensate, i.e., we have a large nucleus moving along the positive longitudinal
direction and colliding with a static or slowly moving quark. This discussion will later
be extended to the fully physical situation of collisions of nuclei in the fragmentation
region of the nucleus.  While this extension is straightforward, the problem of the single charged particle is sufficiently 
involved that it is useful to do it first.

There is an extensive literature on the computation of gluon production in collisions of large 
nuclei at very high energies \cite{kmw}-\cite{mt3}.
The beam fragmentation region can also be studied as a forward limit \cite{gsv,iancu}. Both beam and target then move on the light
cone, but in our case we consider the target to be at rest.

We can understand why the fragmentation region of high energy collisions is distinctively different from the central region.  Let us consider an asymmetric collision between a large particle, which we will call a nucleus and a small particle such as a small nucleus, or a proton or for that matter a quark or gluon.
The current for the sheet of colored glass we shall refer to as that of  the nucleus.  
This current generates a strong field with a characteristic momentum scale associated 
with the average density of charges on it, $Q_{\rm{sat}}$.  If a particle  with transverse momenta 
$k_T$ interacts with the nucleus, it scatters with high probability for $k_T < Q_{\rm sat}$, and above 
this scale the nucleus becomes increasingly transparent.

The saturation momentum scale may be evaluated to be \cite{ Gribov:1984tu, Mueller:1989st}
\begin{equation}
     Q_{\rm sat}^2 = \alpha_s N_c {1 \over {\pi R_A^2}} {{dN} \over {dy}},
\end{equation}
where $\alpha_s$ is the strong coupling constant, $N_c$ is the color factor and $dN/dy$ is the rapidity density of gluons.  This rapidity density should grow like
$e^{\kappa \alpha_sy}$ at large rapidities far from that of the projectile nucleus\footnote{There is only one factor of $\alpha_s$ in this equation
because the scale at which unitarity in scattering sets in involves scattering from all the gluons from the rapidity of interest to that of the projectile, and the integration over rapidity converts $\alpha_s^2 \int ~dy ~ dN/dy \sim \alpha_s dN/dy$, because of the exponentially growing gluon density.  We see that at very high energies the saturation momentum of the nucleus can become very large.}.

Now consider the fragmentation region of the smaller particle.  The saturation momentum of the smaller particle
has not evolved since it is not being evaluated at a rapidity scale far from its own fragmentation region.  On the other hand,
at the fragmentation region of the smaller particle, one is many units of rapidity away from that of the projectile nucleus, and the saturation momentum of the projectile nucleus at asymptotically  high energies evaluated at  these scales can become asymptotically large.  We therefore
have that the saturation momentum projectile and target in the fragmentation region of the target satisfy
\begin{equation}
   Q_{\rm sat}^{\rm targ} \ll Q_{\rm sat}^{\rm proj}.
\end{equation}

It can be shown that the majority of particles are produced in the kinematic region where the produced particle transverse momentum satisfies
\begin{equation}
    Q_{\rm sat}^{\rm targ} \ll p_T \ll Q_{\rm sat}^{\rm proj}
\end{equation}
and that in this region the gluon field of the produced gluon is large enough so that it can be treated classically, but small enough so that
the gluon field equation may be treated as a linear equation \cite{Kovchegov:1998bi}-\cite{dumitru},
\begin{equation}
  1 \ll A_{\rm gluon}^\mu \ll 1/g.
\end{equation}

This observation about the gluon field strength is at the heart of the computation we present here.  We will work to all orders in the strength
of the color field of the nucleus but to lowest order in the field strength of the target particle.  This will allow us to compute the trajectory of the struck target particle, and the induced gluon radiation associated with this collision.

\section{Review of the Properties of a  Color Field of a Sheet of Colored Glass}

Color Glass Condensate refers to an ensemble of classical charge on a sheet at $x^- = 0$.  For an arbitrary four-vector 
$x^{\mu} = (x^+,x^-,x^i)$ we choose light cone coordinates as
\begin{equation}
  x^\pm = {t \pm z \over \sqrt{2}},
\end{equation}
where $z\equiv x_L = (x^+ -x^-)/\sqrt{2}$ is the longitudinal coordinate, and $x^i$ for $i=1,2$ are the transverse coordinates, with
$x_T^2=x^ix^i$. We use the mostly plus metric $g^{+-} =  g^{-+}  = - 1$, $g^{11} = g^{22} = +1$ so that the scalar product of two four-vector is
\begin{equation}
  a \cdot b =  - a^+ b^- -  a^- b^+ + a^i\,b^i. 
\end{equation}
What we gain hereby is that we need not worry about the sign change in transverse components, $a_i=a^i$; 
it is easier to remember the sign in $a_+=-a^-$.
With these conventions, the mass shell constraint $p^2 =  - m^2$ is $p^+ = m_T^2/2p^-$, with the definition $m_T^2=p_T^2+m^2$.

The classical Yang-Mills equations of motion in the presence of an external current $J^{\nu}$ are given by
\begin{equation}
\label{eq:cym}
D_{\mu}F^{\mu\nu} = J^{\nu},
\end{equation}
where the covariant derivative and the field strength are
\begin{eqnarray}
D_\mu & = &  \partial_\mu-igA_\mu, \\
F_{\mu\nu} & = & \frac{i}{g}[D_\mu,D_\nu]=\partial_\mu A_\nu-\partial_\nu A_\mu-ig[A_\mu,A_\nu],
\end{eqnarray}
and $A_{\mu} = A_{\mu}^{a}T_{a}$ are the matrix valued gauge fields. The SU($N_c$) gauge algebra is
\begin{equation}
\left[ T_a,T_b \right]  = if_{abc}T_c, 
\end{equation}
where $f_{abc}$ are the totally antisymmetric structure constants and  $T^a_{bc}=-if_{abc}$ for adjoint representation.

Under a unitary gauge transformation $U(x)$, the gauge field transform as 
\begin{equation}
A_\mu\to A_\mu^\prime=UA_\mu U^\dagger + \frac{i}{g}U\partial_\mu U^\dagger,
\end{equation}
or, equivalently,
\begin{equation}
D_\mu\to D_\mu^\prime = UD_\mu U^\dagger.
\end{equation}
The field strength and the current transform covariantly:
\begin{eqnarray}
F_{\mu\nu} & \to & F_{\mu\nu}^\prime=UF_{\mu\nu} U^\dagger, \\
J_{\mu} & \to & J_{\mu}^\prime=U J_{\mu} U^\dagger. 
\end{eqnarray}
We often also use a matrix-vector notation for transformations of a color vector $F_a,\,a=1,...,N_c^2-1$. If we have
a $d_R\times d_R$ dimensional representation of the generators $T_a$, normalised by ${\rm Tr}~T_aT_b=T_R\delta_{ab}$,
then a matrix representation of the transformation $F^\prime =UFU^\dagger$ can as well be written as
$F^\prime=VF$ or in component form
\begin{equation}
F^\prime_a={1\over T_R}{\rm Tr}(T_aUT_bU^\dagger)F_b=V_{ab}F_b.
\label{V}
\end{equation}
Thus $V$ is an adjoint matrix and $T_a,\,U$ are $d_R\times d_R$ dimensional matrices.

The sheet of Colored Glass may for many purposes be treated as an infinitesimally thin sheet, with color charges
\begin{equation}
  \rho_a(x^-, x_T) = \delta(x^-) \rho_a(x_T)
  \label{sheet}
\end{equation}
on the sheet. A crucial assumption here is that there is no $x^+$ dependence in Eq.~(\ref{sheet}). The physical basis for this is time dilatation,
the fast degrees of freedom are effectively frozen. 
In some circumstances it is useful to spread the charge in Eq.~(\ref{sheet}) out over an interval 
\begin{equation}
0 < x^- < x^-_0,
\end{equation}
where $x^-_0$ is assumed to be very small.  This regularizes our computations and may be thought of as arising from the rapidity 
distribution of gluons between that of the fragmentation region of the nucleus and that where the gluon distribution is measured.  
The spatial rapidity is therefore spread over a finite interval. The color current associated with this source is taken to be
\begin{equation}
  J^\mu_a(x^-,x^i) = \delta^{\mu +}\rho_a(x^-,x^i).
  \label{jplus}
\end{equation}

When computing physical quantities one computes fields in the presence of the sources, and then averages over the sources.  When
the distribution of the  sources is chosen to be Gaussian, this is the McLerran-Venugopalan model \cite{mv}.

There are two gauges, both subclasses of the $A^-=0$ gauge, in which the solution of the classical 
Yang-Mills equations with the current Eq.~(\ref{jplus}) is usually discussed, either $A^\mu=(A^+(x^-,x^i),0,0,0)$ or $A^\mu=(0,0,A^i(x^-,x^i))$ (for reviews, see \cite{ilm,gv}). The solution for the field corresponding to the nucleus is most easily found in the gauge there the field is entirely in the $+$ direction.
Since there is no $x^+$ dependence in the current Eq.~(\ref{jplus}), the only non-zero components of $F^{\mu\nu}$ are
\begin{equation}
    F^{i+}_a = - F^{+i}_a = \partial^i A^+_a
\end{equation}
and the equations of motion reduce to
\begin{equation}
    \partial_T^2 A^+_a = \rho_a(x^-, x^i).
\end{equation}
From this one can then gauge transform $A^+$ to zero, which then generates a transverse $A^i(x^-,x^k)$ where the only 
non-zero components of the field strength are now
\begin{equation}
    F^{i+}_a = - F^{+i}_a = \partial_{-} A^i_a.
\end{equation}
Usually this transverse field is further chosen so that it is zero before the sheet at $x^-<0$ and non-zero after it, for $x_0<x^-$. Asymptotically
this sheet  is essentially a step function $\theta(x^-)$ that is a nonzero constant for $x^- > 0$ (see Fig. 12 of \cite{ilm}).

In what follows, we will actually find it convenient to use a gauge in which $A_i\sim \theta(-x^-)$, i.e., vanishes after
the sheet.  The geometry of the process we are studying is shown in Fig.~{\ref{excitation}.
A particle initially at rest at $(x_L=0, x_T=0)$ collides with a sheet of colored glass moving along $x^-=0$ with the current Eq.~(\ref{jplus})
and we are interested in the radiation field produced. Thus we need the fields far in the future so that, for simplicity, 
it is convenient to have the field of the nucleus vanish after the collision $x^- > 0$.  
Because the collision is singular at $x^- = 0$, the choice where the gauge
field is entirely $A^+$ is perhaps not optimal either (though it can be used to study similar processes, \cite{casalderrey}-\cite{mt3}).
We therefore will work in the gauge for the scattering problem where the field is
entirely a two dimensional transverse field for $x^- < 0$, $A^i_a(x^-,x^k)$ approximately constant as a function of $x^-$
and vanishes for $x^->0$. 

In this gauge, because the single particle 
source is at rest before the collision, we have that the single particle source is not rotated in the background field of the nucleus, 
\begin{equation}
D_{\mu}J^{\mu} = \partial_{\mu}J^{\mu} - ig\left[A_{\mu},J^{\mu} \right] = \partial_+J^{+} = 0.
\end{equation}
Thus the single particle source before the collision is time independent. 

Let us then construct explicitly the gauge rotation that connects the $A^+$ gauge to the $A^i$ gauge.  To transform
$A^+$ to zero the gauge rotation matrix $U$ must solve
\begin{equation} 
    U (\partial^+  - igA^+ ) U^\dagger = 0
\end{equation}
or
\begin{equation}
      \partial_-U^\dagger =- igA^+ U^\dagger.
\end{equation}
A solution where the matrix $U$ is one for $x^-$ positive and outside the thin sheet at $0<x^-<x_0$ for 
$x_0\to0$ is
\begin{equation}
  U(x^-,x^i) = P \exp \biggl \{-ig\int_{x^-}^{\infty} dy^- A^+(y^-, x^i)  \biggr \}.
  \label{U}
\end{equation}
where $P$ is the path-ordering operator. When $x^-<0$, one gets the entire contribution of the sheet and $U$ is constant in $x^-$.
The two dimensional gauge field associated with this rotation is
\begin{equation}
   A^j = {i\over g} U \partial^j U^\dagger
   \label{Ai}
\end{equation}
and, as constructed, is non-zero and constant at $x^- < 0$ and vanishes for $x^->x_0\to0$.

\begin{figure}[!t]
\begin{center}
\includegraphics[width=0.5\textwidth]{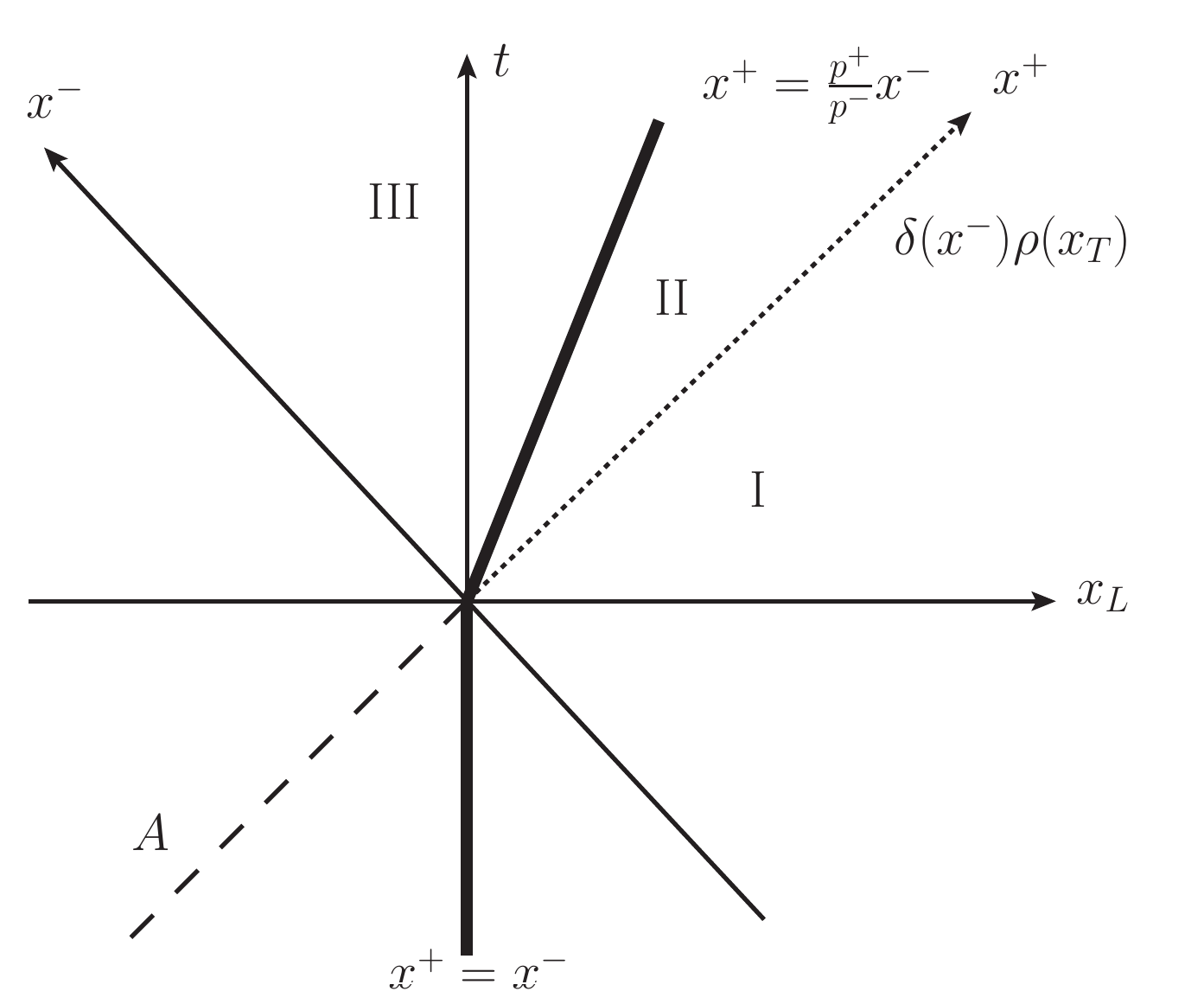}
\caption{\small  Longitudinal kinematics of the process. Nucleus A moving along the light cone with color current $\delta^{\mu +}
\delta(x^-)\rho(x_T)$ meets with a stationary quark and is excited into a classical background Yang-Mills configuration 
$A^\mu=(0,0,A^i \,\,\theta(-x^-))$. In region I, for $x^-<0$, the background field is vacuum equivalent transverse field
\protect(\ref{Ai}) and the fluctuation field $\delta A_i\equiv a_i$ is a gauge rotated Coulomb field of the stationary 
quark \protect(\ref{Ua}). In II and III there is no background field and the fluctuation field contains electrodynamics-like
radiation caused by quark acceleration \protect(\ref{mult1}) and radiation from the quark-nucleus interaction 
\protect(\ref{finres}).
}\label{excitation}
\end{center}
\end{figure}

\section{Field of an Isolated Particle and Wong's Equations}

The single particle is at rest before the collision.  This is because the field strength $F^{\mu  \nu}$
vanishes except in the thin sheet of colored glass.  When it hits the Colored Glass it is accelerated, and begins to radiate.
After the collision, it has a constant velocity to leading order in the strength of the small radiated field.  We want to construct both the radiation field and the motion of the charged particle.

In the construction of the field of the charged particle, the extended current conservation law can cause a potential problem associated with the induced field
of the classical particle rotating the current of the sheet of colored charge.  We deal with this by working in the gauge for the 
small fluctuation field
\begin{equation}
    \delta A^- \equiv a^-= 0
\end{equation}
In this gauge, combined with our choice for the gauge of the background field of the nucleus discussed above, the charges of both the nucleus
and the single particle do not precess. This is since the only non-zero precession term is of order
the single particle field times the source of the single particle field, and this is second order in the strength of the single
particle field and we work only to first order. Note that one might be worried about some rotation induced while the charged 
particle traverses the sheet, but this occurs in a short time, and in our gauge, none of the fields diverge in the limit of 
infinitesimal sheet, and it therefore induces insignificant rotation.

So with these concepts in mind, let us now compute the trajectory of the single color charged particle as it traverses the sheet, determined by Wong's equations \cite{wong,wong2}. The classical particle in a colored field has a classical color vector $T_a(\tau)$, a trajectory 
$x^{\mu}(\tau)$ and momentum $p^{\mu} (\tau)$.  Here $u^{\mu} = dx^{\mu} /d\tau$ and $p^{\mu} = m u^{\mu}$.
The equation of motion is
\begin{equation}
  {{dp^\mu} \over {d\tau}} = gT \cdot F^{\mu \nu} u_{\nu}\equiv gT_a F^{\mu\nu}u_\nu.
\label{wong}
\end{equation}
Due to the asymmetry of $F^{\mu\nu}$ this explicitly conserves the square of the particle four-momentum $p$, i.e. $d(p_\mu p^\mu)/d\tau=0$. 
By demanding covariant conservation $D^\mu J_\mu=0$ of the current
\begin{equation}
  J_\mu(x)=\int d\tau \,T(\tau)u_\mu(\tau)\delta^{(4)}(x^\mu-x^\mu(\tau))
\end{equation}
one sees that the equation for the precession of the colored charge matrix $T$ is
\begin{equation}
  {{dT} \over {d\tau}} = -ig [T, u \cdot A],
\end{equation}
or, in component form
\begin{equation}
  {{dT_a} \over {d\tau}} = -gf_{abc}T^cA_\mu^b u^\mu.
\end{equation}
A formal solution of this first-order matrix equation is the path ordered adjoint exponential
\begin{equation}
   T(x) = P\exp \left (ig\int^x_0 dx_\mu A^\mu\right ) T(0),
\label{gaugerotation}
\end{equation}
where the path ordered integration over $x$ is over the trajectory of the particle.  

Now before the collision, the integration above is entirely time-like, and the fields are 2-d transverse, so the integration over this part 
of the path vanishes, $dx_i A^i=0$.  Near $x^- = 0$, there is the integration across the sheet, but in our gauge the vector 
potential of the background field is so mildly singular, $A_i\sim \theta(-x^-)$, 
that there is no contribution in the limit the sheet shrinks to zero.  
We ignore the contribution of the induced field produced by the single charged particle.  Similarly, after the collision 
there is no contribution, since the background field from the nucleus vanishes there.  Therefore, within the approximation 
of infinitely thin sheet we consider, the color of the particle
does not precess.

Consider then the motion of the particle as given by Wong's equations Eq.~(\ref{wong}). Since in our gauge
$F^{-i}=F^{-+}=0$, the simplest equation is the one with $\mu=-$:
\begin{equation}
{dp^-\over d\tau}=0,
\end{equation}
so that $p^-/m=u^-=dx^-/d\tau$ is a constant, $x^-(\tau)=u^-\tau$ and $\tau$ is effectively
the same as $x^-$. Further, we have $F^{i+}=F_{-i}=\partial_- A^i$ and $u_+=-u^-$ so that the $\mu=i$ 
equation gives
\begin{equation}
{dp^i\over d\tau}=u^-{dp^i\over dx^-} = -gT\cdot (\partial_- A^i)u^-.
\label{sheetkick}
\end{equation}

In this equation, the vector potential is evaluated at $ x_T = 0$, and since $T$ is the color charge at $x_T = 0$, 
the overall result is gauge invariant. Before the sheet at $x^-<0$ $A^i$ is nonzero but effectively constant so that $p^i$ is constant. After the sheet at $x^->0$ $A^i=0$ so that again $p^i$ is constant. 
Taking $p^i=0$ before the sheet and integrating Eq.~(\ref{sheetkick}) across $x^-=0$ gives the particle a kick of magnitude
\begin{equation}
p^i= -g T \cdot {\rm disc} A^i.       
\end{equation}
Finally, Wong's equations explicitly conserve the mass shell condition $p^2=-m^2$ so that
the $\mu=+$ equation simply enforces the mass shell condition.

We expect that the magnitude of the final transverse momentum into which the static quark is scattered is of the order of the relevant dynamical scale,
the nuclear saturation momentum, $p_ip_i=p_T^2\sim Q_{\rm sat}^2$. A rough estimate could be 
$p_T^2\sim \langle\partial_iU\partial_iU^\dagger\rangle\sim Q_{\rm sat}^2\log(1/(\Lambda_{\rm QCD} r),$ where $r$ is the
transverse separation of the two $U(x_T)$ matrices and we used a standard formula \cite{ilm} for the $UU^\dagger$ correlator.
This expression diverges in the limit that we take the size of the particle probing our system to zero.  This is an artifact of
the high momentum divergence for average transverse momentum squared associated with the $1/p_T^4$ behavior
of large angle scattering.

The current for the single particle may be computed from the constant trajectory 
$u^\mu =dx^\mu/d\tau=p^\mu/m$ and  is for $x^- > 0$ specified as\footnote{As argued above, for asymptotically
thin sheet the color matrix $T$ does not rotate.}
\begin{eqnarray}
  J^\mu(x) &=&T \int d\tau u^{\mu} \delta^{(4)} (x^\mu - u^{\mu} \tau)\,\theta(x^-)\\
  &=&Tu^\mu \int d\tau \delta(x^--u^-\tau)\delta(x^+-u^+\tau)\delta^{(2)}(x^i-u^i\tau)\,\theta(x^-)\\
  &=&T{p^\mu\over p^-}\delta(x^+ -{p^+\over p^-}x^-)\delta^{(2)}(x^i-{p^i\over p^-}x^-)\,\theta(x^-).
\label{current}
\end{eqnarray}
Notice that the delta function of involving $x^-$ simply sets the momentum space rapidity of the scattered particle 
equal to the coordinate space rapidity. In momentum space this current is
 \begin{equation}
J^\mu(k)=\int d^4x\, e^{i(k^+x^-+k^-x^+-k^ix^i)}J^\mu(x)={Tp^\mu\over i(p\cdot k-i\epsilon)}.
\label{curr}
\end{equation}
Similar expressions are valid for $x^-<0$; the sign of Eq.~(\ref{curr}) as well as the sign of $\epsilon$ are then
changed.

\section{The Coulomb Potential in $\delta A^- = 0$ Gauge}

We have chosen the background color field of the nucleus to be of the form $A^\mu=(0,0,A^i)$. In general, when discussing
the total nucleus-quark system we will use the $A^-=0$ gauge. We will write the total field in the form
$A^\mu+\delta A^\mu\equiv A^\mu+a^\mu$, where $\delta A^\mu=a^\mu$ is of lower order.

Before scattering off the sheet of colored class, the field of our classical particle is Coulombic.    We have,
with $r=\sqrt{x_L^2+x_T^2}= \sqrt{(x^+ - x^-)^2/2 + x_T^2}$, 
\begin{equation}
    a^\mu_{Coul} = {1  \over {4\pi r}} \delta^{\mu 0},\qquad a^+= a^- ={ 1\over {\sqrt{2}}} {1 \over {4 \pi r}},
\end{equation}
where the matrix $g T$ that multiplies the charge has been suppressed.
This field is shifted into the light cone gauge $a^- = 0$ by an infinitesimal gauge
transformation $U=e^{ig\Lambda}\approx 1+ig\Lambda$:
\begin{equation}
   a^\mu\to a^\mu+D^\mu(A)\Lambda,\quad D^\mu=(\partial^+,\partial^-,\partial^i-igA^i),
\end{equation}
or, in component form,
\begin{eqnarray}
 a^- &\to& { 1\over {\sqrt{2}}} {1 \over {4 \pi r}} + \partial^- \Lambda = 0,\label{deltaAm}\\
   a^+ &\to&  { 1\over {\sqrt{2}}} {1 \over {4 \pi r}} + \partial^+ \Lambda ,\\
   a^i &\to& 0 + D^i \Lambda .
\end{eqnarray}
We are actually only concerned with this field for $x^- < 0$, since the field for $x^+ > 0$ will include the radiation field 
and will be determined by solving a boundary value problem at $x^- = 0$ with boundary conditions determined by the field for $x^- <0$.

The field $\Lambda $ (related to the infinitesimal transformation) is determined from Eq.~(\ref{deltaAm}):
\begin{equation}
  \Lambda =- \int ~dx^- {1 \over {4\pi \sqrt{2} }} {1 \over \sqrt{(x^+ - x^-)^2/2 + x_T^2}}.
\end{equation}
This gives
\begin{equation}
  \Lambda = {1 \over {4\pi }} \ln{\left[ \frac{1}{x_T} \left((x^+ - x^-)/\sqrt{2} + \sqrt{ (x^+ -x^-)^2/2 + x_T^2 } \right) \right]}.
\end{equation}
Note that $\partial^-\Lambda=-\partial_+\Lambda=\partial_-\Lambda$, where the first equality is general for LC coordinates, and the second holds since
$\Lambda$ depends on the difference $x^+-x^-$. 
This yields for the $+$ component
\begin{equation}
  a^+ =  {\sqrt{2} \over {4 \pi r}}.
\end{equation}

Concerning the $i$ component, consider first the vacuum case so that $D_i=\partial_i$. Then 
\begin{equation}
  a^i =\partial_i\Lambda=- { 1 \over {4\pi}}{ x^i \over x_T^2} {x_L \over r}={1\over\partial^-}{x^i\over 4\pi\sqrt2\, r^3}={1\over\partial_+}
  \partial^i{1\over 4\pi\sqrt2\, r}.
  \label{varia}
\end{equation}
We have included some explicit relations which come in handy later. These fluctuation fields satisfy further 
\begin{equation}
\partial_+a^+ +\partial_i  a^i=0,\quad\bigl( -2\partial_+\partial_-+\partial_i\partial_i\bigr)a^i=0, \quad
\partial_i a_j-\partial_j a_i=0.
\end{equation}
The last equation means that there is no longitudinal magnetic field.

One may ask whether there is residual gauge freedom in the transverse fields $a_i$. In fact, one can
still do a U(1) gauge transformation ($C$ = constant)
\begin{equation}
 a_i\to a_i+\partial_i\chi,\quad \chi = C\log x_T,\quad \partial_i\chi=C{x_i\over x_T^2},\quad \partial_T^2\chi=0.
\end{equation}
This transformation is the same as
\begin{equation}
 \Lambda \to \Lambda + C\log x_T
\end{equation}
and the 2-divergence of $a_i$ is invariant under it
\begin{equation}
 \partial_i a^i = {x_L\over 4\pi r^3} \to \partial_i (a^i+\partial_i\chi)=\partial_i a^i. 
\end{equation}

For the vacuum case, no nuclear background field, we now have the full set $a^\mu=(a^+,0,a^i)$. 
One can check that this form reproduces the desired structure for $F^{\mu \nu}$ as it must.

We will later need some Fourier transformations of the vacuum fields along the line $x^-=0$.
To summarise:
\begin{equation}
\hspace{-2cm}
\int_{-\infty}^{+\infty} dx^+ d^2k_T\, e^{+i(k^-x^+ - k^ix^i)}{1\over 4\pi r}\biggl(1,{x^i\over r^2},{x_Lx^i\over x_T^2}\biggr)=
{\sqrt2\over k_T^2+2(k^-)^2}\biggl(1,ik^i,-{k^i\over \sqrt2 k^-}\biggr).
\label{fouriers}
\end{equation}

\section{Field of test quark in a strong background field}
For the non-abelian problem, one has to generalize the Coulomb solution to the case where for $x^- < 0$ there is a strong background field 
$-igA_i=U\partial_i U^\dagger$, a two dimensional gauge transform of vacuum.  Here $U$ is the gauge transformation matrix Eq.~(\ref{U}) transforming from
the $A^+$ gauge to the $A^i$ gauge. 

The fluctuation equation for the transverse field $a^i$ in the
$A^i$ gauge is (see Eq.~(\ref{aieq}) below)
\begin{equation}
 D_\mu D^\mu a^i=(-2\partial_+\partial_-+D_iD_i)a^i=-2\partial_+\partial_- a^i+U\partial_T^2(U^\dagger a^i)=Uj^i,
\end{equation}
where $j^i$ is a current representing a color source at $x_L=x_T=0$ in the $A^+$ gauge, in which there is no
transverse background field. Multiplying from the left by $U^\dagger$ 
and noting that $\partial_-U=0$ (i.e.there is no $x^+$ dependence in the matrix $U$)  
shows that $U^\dagger a^i$ satisfies vacuum equations:
\begin{equation}
(-2\partial_+\partial_- +\partial_T^2)U^\dagger a^i=j^i,
\end{equation}
i.e.,
\begin{equation}
a_i(x^+,x^-,x^k) = U(x_T)a_i^{\rm vac}(x^+,x^-,x^k)\theta(-x^-).
\label{Ua}
\end{equation}
In Eq.~(\ref{Ua})  $a_i^{\rm vac}$ satisfies the vacuum equations (see previous Section) and we inserted a theta function to remind that
this discussion is relevant at $x^-<0$. The vacuum fields $a_i^{\rm vac}$ are determined with the color vector
$T$ (sometimes $g$ is also appended) in the current.

One consequence of the appearance of $U$ in Eq.~(\ref{Ua}) is that the 2-divergence is modified by a coupling to the
background field $A_i$:
\begin{equation}
\partial_ia_i =(\partial_i U(x_T))a_i^{\rm vac}+U\partial_i a_i^{\rm vac}=igA_i a_i + U(x_T){ T x_L\over 4\pi r^3},
\label{partiala}
\end{equation}
where we inserted $\partial_i U=igA_i U$ and restored $T$.  Also a longitudinal magnetic field is generated:
\begin{equation}
\epsilon_{Lij}\partial_ia_j =\epsilon_{ij}(\partial_i U) a_j^{\rm vac}=ig\epsilon_{ij}A_i a_j.
\label{Bl}
\end{equation}
These equations are important since they give the first derivatives of the transverse radiation field,
which are needed to solve the radiation equations, see Section \ref{res} below. We also remind that the
$a$th color component of the vector $igAa$ is $(igAa)_a= gf_{abc}A^ba^c$.

In a strong background field the Coulomb solution for $x^- < 0$ is then
\begin{equation}
            a^0_{Coul} = {1 \over {4 \pi r}} U(x_T)T.
\end{equation}
We could redefine $T$ by an overall rotation to give
\begin{equation}
  \overline{T} = U^\dagger(0) T 
  \label{tbar}
\end{equation}
so that the Coulomb potential at $x_T=0$, where the quark is sitting,  is $T/(4\pi r)$ without any rotation.
However, the choice $U(0)=1$ is possible and will be made. The field in the gauge $a^- = 0$ is as before obtained by a small gauge transformation with
\begin{equation}
\partial_- \Lambda=-{1\over \sqrt2\, 4\pi r}U(x_T)T
\end{equation}
so that
\begin{equation}
  a^+ = {1 \over {\sqrt{2}}} a^0_{Coul} - \partial_- \Lambda = \sqrt{2} a_{Coul}^0
\end{equation}
and
\begin{equation}
  \hspace{-1cm}  a^i = D^i\Lambda=U\partial_i(U^\dagger\Lambda)=
     - {1 \over {\sqrt{2}}} D^i {1 \over \partial^-} a_{Coul}^0 = -{1 \over {\sqrt{2}}} U {1\over \partial^-} 
   \partial^i {T \over {4\pi r}} .
\end{equation}
This, of course, is the same as Eq.~(\ref{Ua}), see also Eq.~({\ref{varia}).

\section{Radiation from a Point Particle Crossing a Sheet, QED and rapidity}\label{edlike}

In this section we review the computation of the radiation from a  charged electromagnetic particle getting an impulse kick at
$x^- = 0$.  This review will make the discussion of the non-abelian problem more transparent.

Let us assume we have a particle that is at rest at the origin for $t < 0$ and is spontaneously accelerated to a
particle with constant momentum $p^\mu$ at $t = 0$.  The current is
\begin{equation}
   J^\mu = e \left[\delta^{\mu 0} \int_{-\infty}^0~  d\tau \,\,\delta(t-\tau)\delta^{(3)}(\vec{x}) +u^\mu\int_0^{\infty}~ d\tau  \,\,
   \delta^{(4)} ( x^\mu - u^\mu \tau) \right].
\label{JED}
\end{equation}
Here $u^\mu = p^\mu/m$ is the four velocity of the particle after the collision. In light cone gauge $A^- = 0$ the solution for the vector potential is
\begin{equation}
   A^+ = {1 \over \partial^-} (\nabla  \cdot A + {1 \over \partial^-} J^-)
\end{equation}
and 
\begin{equation}
  \partial_\mu\partial^\mu A^i = J^i - {\partial^i \over \partial^-} J^-.
\end{equation}
If we rewrite this in Fourier space, then the distribution of radiation is 
\begin{equation}
\hspace{-2cm}16\pi^3 k {{dN} \over {d^3k}} = 
\lim_{k^2 \rightarrow 0}~k^4  A^i(k) A^i(-k)= \left [J^i(k) - {k^i \over k^-} J^-(k) \right ] \left [J^i(-k) - {k^i \over k^-} J^-(-k) \right ].
\label{mult1}
\end{equation} 
In the following, we will assume $k^2 = 0$. Up to a term that vanishes when $k^2 = 0$, upon use of
current conservation $k \cdot J = 0$, the right hand side is algebraically identical to
\begin{equation}
16\pi^3  k {{dN} \over {d^3k}} = J_\mu(k)J^\mu(-k)
\label{mult2}
\end{equation}
which is the ordinary text book expression.

We can now compute the Fourier transform of the current as
\begin{equation}
   J^\mu = e\left[ \delta^{\mu 0} {1 \over {ik^0}} - u^\mu {1 \over (i k\cdot u)} \right].
\label{EDsol1}
\end{equation}
After a little algebra, we find that
\begin{equation}
    J^2 = e^2 v^2 {1 \over k^2} {{\sin^2\theta} \over {(1 - v \cos\theta)^2} },
\end{equation}
where the angle $\theta$ is between that of the 3-dimensional velocity vector ${\bf v}={\bf p}/E$ of the particle and the emitted photon
$(k= \vert{\bf k}\vert=\omega)$.

Although we are mainly interested in a stationary initial quark, it may be useful to give some expressions
for a more general radiation process with momenta $q\to k+p$ and write them in terms of rapidities,
\begin{equation}
\hspace{-2cm} 
k^\mu=k_T({\textstyle{{1\over\sqrt2}}}  e^y,{\textstyle{ {1\over\sqrt2}}} e^{-y},\cos\phi, \sin\phi),\quad
q^\mu=({\textstyle{{m_{Tq}\over\sqrt2}}}  e^{y_q},{\textstyle{ {m_{Tq}\over\sqrt2}}} e^{-y_q},q_T\cos\phi_q,q_T\sin\phi_q),
\end{equation}
where $m_{Tq}^2=m^2+q_T^2$, and similarly for $p^\mu$.
The radiation current in Eqs.~(\ref{curr},\ref{EDsol1}) for instantaneous acceleration $q\to p$ at $x^-=0$ is
\begin{equation}
J^\mu(k)=+iT\biggl({q^\mu\over q\cdot k}-{p^\mu\over p\cdot k}\biggr).
\end{equation}
Using the kinematic relation $|k^i-k^-p^i/p^-|^2=-2(k^-/p^-)\,p\cdot k$  this gives
\begin{equation}
\hspace{-1cm} -k^2A_i(k)=
J^i(k)-{k^i\over k^-}J^-(k) =2iT\left[{k^i-{k^-\over q^-}q^i \over |k^i-{k^-\over q^-}q^i|^2}
-{k^i-{k^-\over p^-}p^i \over |k^i-{k^-\over p^-}p^i|^2}\right],
\label{Jeff}
\end{equation}
where the first term corresponds to $\theta(-x^-)$ and the second to $\theta(x^-)$.
The ratio $k^-/p^-$ is the fractional light cone energy taken by the photon from the emitting charge.
The multiplicity can now be computed from Eq.~(\ref{mult1}) or Eq.~(\ref{mult2}). For a massless quark one has 
\begin{equation}
   \hspace{-2cm}{dN\over dyd^2k_T}=
   {g^2T_aT_a\over (2\pi)^3}\,{1\over k_T^2}{\cosh(y_p-y_q)-\cos(\phi_p-\phi_q)\over(\cosh(y-y_q)-\cos(\phi-\phi_q)) (\cosh(y-y_p)-\cos(\phi-\phi_p))}.
\end{equation}
Integration over the azimuthal angle of the produced gluon produces $dN/dydk_T^2$. This  can also be done analytically using
\begin{equation}
\hspace{-2cm}  \int_0^{2\pi}{d\phi\over (\cosh y_1-\cos(\phi-\phi_1) (\cosh y_2-\cos(\phi-\phi_2)}={2\pi(\coth y_1+
    \coth y_2)\over \cosh(y_1+y_2)-\cos(\phi_1-\phi_2)}.
\end{equation}

Initial static quark, the case were interested in, is obtained from the general formulas in the limit of large mass, 
$m\gg q_T,\,p_T$ and $y_q=0$. If the static quark is accelerated into rapidity $y_p$, the gluon distribution is
\begin{equation}
    {dN\over dyd^2k_T}={g^2T_aT_a\over (2\pi)^3}\,{1\over 2k_T^2}\biggl({\sinh(y_p)\over \cosh(y-y_p)\cosh(y)}\biggr)^2.
    \label{glrap}
\end{equation}
The distribution, plotted in Fig.~\ref{N(y)} is symmetric around $y=y_p/2$ and has a broad plateau around the
maximum. The maximum value is $4\tanh^2{y_p\over 2}$ and the value at $y=0$ or $y=y_p$ is $\tanh^2 y_p$.
For large $y_p$ and small $y$, in the ``target fragmentation region'', the distribution grows like
\begin{equation}
\biggl({2\over 1+e^{-2y}}\biggr)^2.
\label{fragrad}
\end{equation}

\begin{figure}[!t]
\begin{center}
\includegraphics[width=0.5\textwidth]{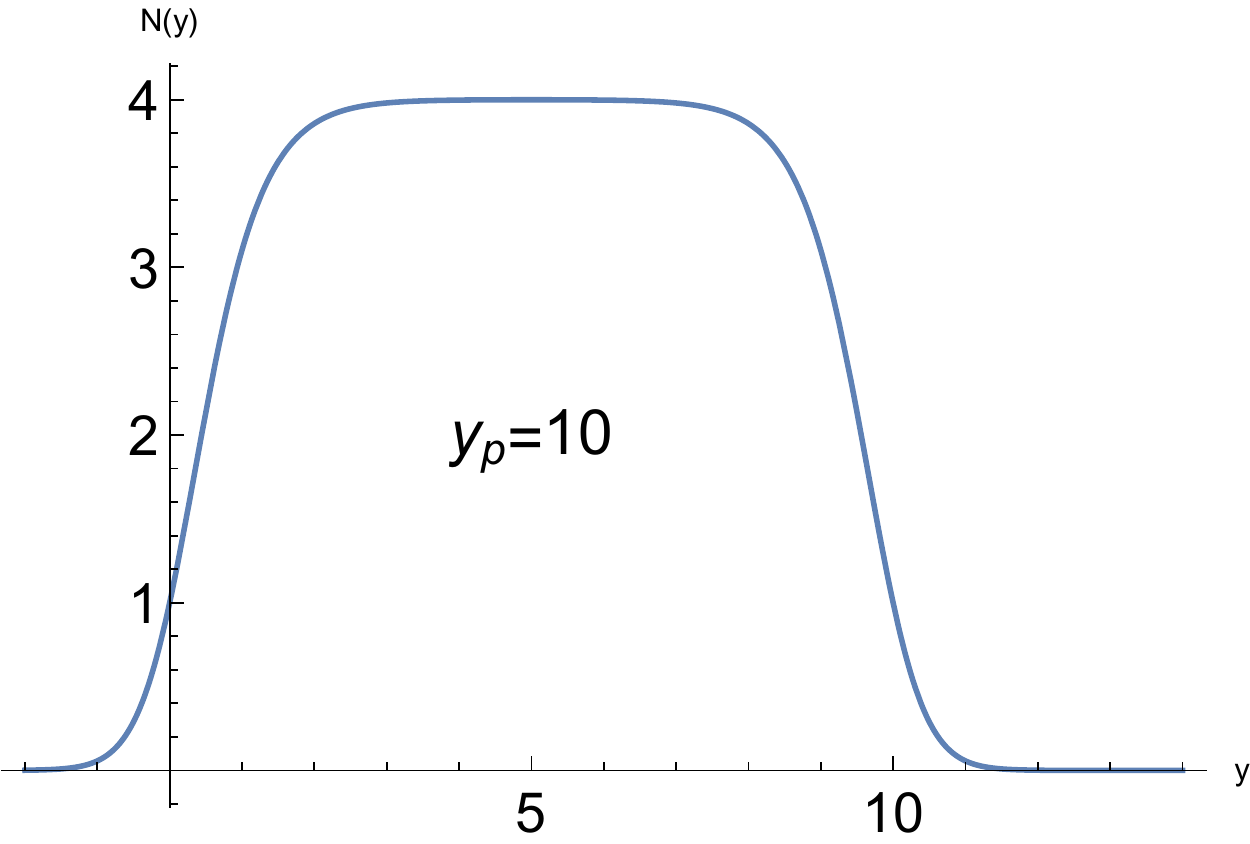}
\caption{\small  Rapidity distribution (bracketed factor in Eq.~\protect(\ref{glrap}) is plotted) of gluons 
emitted in an acceleration of a static quark to rapidity $y_p=10$ by the
nucleus. The maximum is at $y_p/2=5$ and the curve is symmetri around the maximum value.}
\label{N(y)}
\end{center}
\end{figure}

\section{Boundary Conditions}
We have now discussed the transverse fluctuation field $a_i$ at $x^->0$ before the arrival of the nucleus as well 
as the classical radiation from the acceleration of the quark. What is missing is $a_i$  after the nucleus has passed.
However, we have fixed the gauge so that there is no background field at $x^->0$. The solution thus is a free
plane wave and one only needs the boundary conditions at $x^-=0$.
 
 Actually we already know the boundary condition for $a_i$ when approaching from the direction of $x^-<0$;
 according to Eq.~(\ref{Ua}) the value on the surface $x^-=0$ is
 \begin{equation}
   a_i^S(x^+,x^k)\equiv a_i(x^+,x^-=0,x^k)=U(x_T)a_i^{\rm vac}(x^+,x^-=0,x^k).
   \label{bdry}
\end{equation}
But are there discontinuities on the nuclear sheet at $x^-=0$? By studying fluctuation equations we shall show
that there is a discontinuity in $a^+$ but not in $a^-$ or $a^i$.

The equation for the small fluctuation field is
\begin{equation}
  D^2 a^\mu - D^\mu (D \cdot a) - 2ig F^{\mu \nu}a_\nu = Uj^{\mu}.
\end{equation}
The only place where one might generate a discontinuity of the field $a$ is when a derivative with respect to  $x^-$ multiplies a field $A^i$.  
This only occurs in the plus component of the equation of motion:
\begin{equation}
D^2a^+ - D^+(D \cdot a) -2igF^{+i}a_i = Uj^+,
\end{equation}
which can be rewritten as
\begin{equation}
U\partial^2_T(U^{\dagger}a^+) + \partial_- \left (-\partial_+a^++D_ia^i+2igA^i a_i \right ) -2igA^i\partial_- a^i = Uj^+.
\end{equation}
Collecting all the terms  containing $\partial_-$ in the $+$ equation one sees that they are 
\begin{equation}
  \partial_-(-\partial_+a^++D_ia^i+2igA^i a_i)-2igA^i\partial_- a^i.
\end{equation}
The discontinuity in $D_ia^i$ combines with that in the next term and  any discontinuity in $A^i$ can be cancelled by a discontinuity in $\partial_+a^+$:
\begin{equation}
 {\rm disc} \, \partial_+a^+= ig~ {\rm disc}(A^i)  a_i.
\end{equation}
In the $i$ component, there is no such term (in the gauge $a^-=0$ in which we work).  This equation forces $a^i$ to be continuous.

The radiation may be computed by knowing the asymptotic behaviour of the field $a^i$, which satisfies the fluctuation equation
\begin{equation}
D^2a^i - D^i(D\cdot a) = Uj^i.
\label{aieq}
\end{equation}
For $x^- > 0$,  where the matrix $U=1$, this equation simplifies to 
\begin{equation}
\partial^2a^i - \partial^i(D\cdot a) = j^i.
\end{equation}
Further, using the equation for the minus component of the current,
\begin{equation}
 \partial^- (D \cdot a) = - j^-
\end{equation}
we find that for $x^->0$ the equation for $a_i$ simply is
\begin{equation}
\partial^2a^i = j^i - \partial^i \frac{1}{\partial^-}j^-.
\end{equation}

\section{Radiation from the quark-nucleus interaction}\label{res}
The transverse radiation field is
\begin{equation}
a_i(x^+,x^-,x^j)=\int {d^4k\over (2\pi)^4} e^{-i(k^+x^-+k^-x^+-k_jx_j)}\,\,a_i(k^-,k^+,k^j)
\label{four}
\end{equation}
and radiation is computed from $k^2 a_i(k^-,k^+,k^j)$. We know that $a_i$ is a free field at $x^->0$ and further
we know its boundary value Eq.~(\ref{bdry}) at $x^-=0$. With this information we can construct the full
solution.

First, from the fact that one has a free solution for $x^->0$ one can conclude that 
\begin{equation}
 a_i(k^-,k^+,k^j)={i\over k^+-{k_T^2\over 2k^-}+i\epsilon}a_i(k^-,k^j)={-2ik^-\over k^2+i\epsilon}a_i(k^-,k^j).
 \label{ai4}
\end{equation}
Inserting this to Eq.~(\ref{four}) gives
\begin{equation}
 a_i(k^-,x^-,k^j)=\theta(x^-)\exp(-i{k_T^2\over 2k^-}x^-)a_i(k^-,k^j),
\end{equation}
where $a_i(k^-,k^j)$ is the Fourier transform of the
boundary value $U(x_T)a_i^{\rm vac}(x^+,0,x^j)$ in Eq.~(\ref{bdry})
\begin{eqnarray}
a_i(k^-,k^j)&=&\int dx^+ d^2x e^{i(k^-x^+-k^jx^j)}\,U(x^j)\,a_i^{\rm vac}(x^+,0,x^j)\\
&=&\int {d^2q\over (2\pi)^2}U(k_j-q_j){-q_i\over k^-(q_T^2+2(k^-)^2)}.
\label{convol}
\end{eqnarray}
The Fourier transformation was taken from Eq.~(\ref{fouriers}). From Eq.~(\ref{ai4})
\begin{equation}
ik^2 a_i(k^-,k^+,k^j)=2k^-a_i(k^-,k^j).
\label{fin2}
\end{equation}
Radiation from the quark-sheet collision is then computed from (color vector $T$ is now explicitly written)
\begin{equation}
16\pi^3   {{dN} \over {dy d^2k}} = \left\langle\left\vert \int {d^2q\over (2\pi)^2}U(k_j-q_j)T{2q^i\over
 q_T^2+2(k^-)^2}\right\vert^2\right\rangle_\rho,
\label{finres}
\end{equation}
which, together with radiation from acceleration of the quark,  is the main result of this article. For
quantitative evaluation one firstly has to compute the convolution in Eq.~(\ref{finres}) and then perform the
quantum average over an ensemble of color distributions.

The computation of the convolution and color ensemble averaging is a challenging task and we shall here only show how
the result approaches the Gunion-Bertsch formula \cite{gb,kmw} for small $p_T$ particle production in the
central region.

Note first that Eq.~(\ref{convol}) contains part of the ED-like radiation solution discussed in Section \ref{edlike}.
To avoid double counting, this should be eliminated. To this end, write
\begin{equation}
a^ i = \beta^i_1 + \beta^i_2.
  \end{equation}
 We let $\beta_1$ be the solution to the free equations of motion in the presence of the current, that is the analog of the 
 electrodynamics problem for all $x^-$:
 \begin{equation}
 \partial_\mu\partial^\mu \, \beta^i_1 = j^i - {{\partial^i} \over \partial ^-} j^-,
 \end{equation}
 where the current is Eq.~(\ref{JED}) with $e$ replaced by  the unrotated color vector $T$.
 The solution is precisely the solution Eq.~(\ref{EDsol1})  to the electrodynamics problem of radiation from a charged particle.
 
 We let $\beta_2$ be the solution of the free zero external current wave equation 
 $ \partial_\mu\partial^\mu \, \beta^i_2 =0$ subject to the boundary condition
 \begin{equation} 
   \beta^i_2 \mid_{x^- = 0} =  U a_i^{\rm vac} \mid_{x^- = 0}  - \beta^i_1\mid_{x^- = 0} 
    =  U a_i^{\rm vac}\mid_{x^- = 0}  - a_i^{\rm vac}\mid_{x^- = 0} 
    \label{subtr}
 \end{equation}
 This solution for $x^- > 0$ will satisfy correct equations of motion with the proper boundary conditions at $x^- = 0$.
 Here one has subtracted the contribution of $\beta^i_1$ at the surface $x^- = 0$ which is simply 
 the Coulomb solution at $x^- = 0$, $a_i^{\rm vac}$,  the solution
 with unrotated charge vector $T$.

To compute the convolution Eq.~(\ref{convol}) we construct a derivative or momentum expansion as follows. Write the boundary or
surface value in the form
\begin{equation}
a_i^S=a_i(x^+,x^-=0,x^j) = U(x_T)a_i^{\rm vac}(x^+,x^-=0,x^j)=\partial_i\eta
+\epsilon_{ij}\partial_j \chi. \label{expans}
\end{equation}
The two functions $\eta,\chi$ can be projected out ($\epsilon_{12}=-\epsilon_{21}=1$)
\begin{equation}
 \partial_T^2\eta = \delta_{ir}\partial_ra_i=\partial_ia_i^S,\quad \partial_T^2\chi=\epsilon_{ir}\partial_r a_i^S.
\end{equation}
Solving $\eta,\chi$ from here and inserting back to Eq.~(\ref{expans}) gives the indentity
\begin{equation}
a_i^S = (\delta_{ij}\delta_{sr}+\epsilon_{ij}\epsilon_{sr}){1\over\partial_T^2}\partial_j\partial_ra_s^S.
\label{ddee}
\end{equation}
In the last term one can write
\begin{eqnarray}
   \partial_ra_s^S&=&\partial_rU\,a_s^{\rm vac}+U\partial_ra_s^{\rm vac}\nonumber\\
   &=&igA_rU a_s^{\rm vac}+U\partial_ra_s^{\rm vac}\nonumber\\
    &=&igA_r a_s^{\rm vac}+\partial_ra_s^{\rm vac}+{\cal O}(U-1).
\end{eqnarray}
Inserting this back to Eq.~(\ref{ddee}) gives the approximation
\begin{equation}
a_i^S = (\delta_{ij}\delta_{sr}+\epsilon_{ij}\epsilon_{sr}){1\over\partial_T^2}\partial_j  (igA_r a_s^{\rm vac})+a_i^{\rm vac}.
\label{ddee2}
\end{equation}
The last term just goes through Eq.~(\ref{ddee}) unchanged, but 
disappears when the inhomogeneous solution is
subtracted as in Eq.~(\ref{subtr}). In Fourier space we thus have
\begin{equation}
  ik^2a_i^S=(\delta_{ij}\delta_{sr}+\epsilon_{ij}\epsilon_{sr})gf_{abc}{k_j\over k_T^2}
  \int {d^2q\over (2\pi)^2}A_r^b(k_m-q_m)a_s^c(k^-,q_m)2k^-.
  \label{fin1}
\end{equation}
Squaring this and summing over $i$ involves the tensorial structure
\begin{equation}
 \sum_{i=1}^2 (\delta_{ij}\delta_{sr}+\epsilon_{ij}\epsilon_{sr})k_jA_ra_s  (\delta_{i\alpha}\delta_{\beta\gamma}+
 \epsilon_{i\alpha}\epsilon_{\beta\gamma})k_\alpha A_\beta a_\gamma=k_T^2A_T^2a_T^2
 \label{tens}
\end{equation}
so that
\begin{equation}
 \hspace{-2cm}{dN\over dy d^2k_T}={g^2\over 16\pi^3}f_{abc}f_{a{\hat b}{\hat c}}{1\over k_T^2}
 \int{d^2q_1 d^2q_2\over (2\pi)^4}A_r^b(k^i-q_1^i)A_r^{\hat b}(k^i-q_2^i) a^c_s a^{\hat c}_s(k^-,q_2^i)(2k^-)^2.
 \label{res1}
\end{equation}

To proceed further one must go beyond the classical approximation by introducing quantum expectation
values \cite{ilm} of the background field correlator
\begin{equation}
 \langle A_i^a(k_1)A_j^b(k_2)\rangle=\delta_{ab}(2\pi)^2\delta^{(2)}(k_1-k_2)g^2\mu^2{k_1^ik_2^j\over k_{1T}^2k_{2T}^2}.
\end{equation}
Fourier transformations of $a_i^{\rm vac}$ along the line $x^-=0$ were listed in Eq.~(\ref{fouriers}). The relevant one is
\begin{equation}
 2k^-a_i^{\rm vac}(k^-,q^j)={-2q^i\over q_T^2+2(k^-)^2},
\end{equation}
the extra factor $k^-$, which arose from constructing $k^2$, cancels. 
For the color factors one can write $f_{abc}f_{ab{\hat c}}T_cT_{\hat c}=N_c T_{\hat c}T_{\hat c}=N_cN_g$.
With these simplifications Eq.~(\ref{res1}) reduces to the form
\begin{equation}
  {dN\over dy d^2k_T}={g^4\mu^2\over 16\pi^3}N_cN_g{1\over k_T^2}\int {d^2q\over (2\pi)^2}{1\over |k^i-q^i|^2}
  {q_T^2\over (q_T^2+2(k^-)^2)^2}.
  \label{res2}
\end{equation}
If $y$ is the gluon rapidity, $2(k^-)^2=k_T^2e^{-2y}$. At large $y$ away from the fragmentation region at $y=0$ this vanishes
and the momentum integral in Eq.~(\ref{res2}) approaches the standard GB form
\begin{equation}
{1\over k_T^2}\int {d^2q\over (2\pi)^2}{1\over |k^i-q^i|^2}
  {1\over q_T^2}.
\end{equation}

Consider then including the last term $a_i^{\rm vac}$ in Eq.~(\ref{ddee2}), which was subtracted 
as part of the inhomogeneous solution. 
Due to the antisymmetric tensor $f_{abc}$ in the
first term, upon squaring the interference term anyway vanishes. The square of this term gives the gluon distribution
\begin{equation}
{dN\over dy d^2k_T}={g^2\over 16\pi^3}T_aT_a{4k_T^2\over (k_T^2+2(k^-)^2)^2}=
{g^2\over 16\pi^3}T_aT_a{1\over k_T^2}\biggl({2\over 1+e^{-2y}}\biggr)^2.
 \label{res3}
\end{equation}
This is precisely the fragmentation region radiation in the limit of large quark mass, computed in Eq.~(\ref{fragrad}).

Our final result for gluon production is obtained by taking $ik^2 a_i$ from Eqs.~(\ref{convol},\ref{fin2}) for the homogeneous 
$\beta_2$ contribution (quark-sheet collision) and from Eq.~(\ref{Jeff}) for the inhomogeneous 
$\beta_1$ contribution (quark acceleration), summing and absolute squaring.
In the above discussion, we have ignored the contribution for a possible interference term.  This term vanishes.  
This follows because the color structure of the sum is $UT + T$ so that the interference is $\sim UT\cdot T$. Taking
ensemble average over $\rho$ distribution leads to $\langle U\rangle=0$ and interference vanishes. In the weak field
approximation Eq.~(\ref{fin1}) the interference term is $\sim f_{abc}T_aT_c=0$ and vanishes on tree level.
Gluon production therefore arises from two non-interfering contributions:  one that is the generalization of the QED radiation 
process (type $\beta_1$) and another which is unique to QCD, and arises 
from the disturbance of a Coulomb field composed of colored gluons that is disturbed during the collision process (type $\beta_2$)

\section{Rapidity and $k_T$ distribution}
Our goal is to study the fragmentation region so we are interested in the rapidity dependence of the 
produced radiation. The ED-like radiation from the acceleration of the quark when crossing the
nuclear sheet was studied in Section \ref{edlike}, see Fig.~\ref{N(y)} for distribution at fixed $k_T$ and
fixed acceleration. This is the inhomogeneous solution of the radiation equation and depends on the
path of the accelerated quark. The homogeneous solution depends on the color charge distribution of
the sheet colliding with the quark and in Eq.~(\ref{res2}). Its rapidity dependence 
is built in $2(k^-)^2=k_T^2e^{-2y}$. To see what it is quantitatively, we have to introduce 
an IR divergence regulator mass $m$ in the momentum integral in  Eq.~(\ref{res2}). It can then be written
in the form
\begin{eqnarray}
&& {1\over k_T^2}\int {d^2q\over (2\pi)^2}{1\over |k^i-q^i|^2+m^2}{q^2\over (q^2+m^2+k_T^2 e^{-2y})^2}\\
&&={1\over 4\pi k_T^4}\int_0^\infty du{u\over (u+\hat M^2)^2}{1\over\sqrt{(1-u)^2+2\hat m^2(1+u)+\hat m^4}}\nonumber\\
&&={1\over 2\pi k_T^4}\biggl[\log{k_T\over m \sqrt {m^2/k_T^2+e^{-2y}}}-{1\over2}+{\cal O}\biggl({m^2\over k_T^2}\biggr)
\biggr],\label{sheety}
\end{eqnarray}
where $\hat M^2=m^2/k_T^2+e^{-2y}$, $\hat m=m/k_T$. The integral can be done in closed form and expressed
rather compactly by introducing $v^2=1+2\hat m^2+2\hat M^2+(\hat M^2-\hat m^2)^2$. What is relevant is
the limit $m\ll k_T$ shown in Eq.~(\ref{sheety}). The result has the characteristic features:
\begin{itemize}
\item The $k_T$ distribution of the quark-sheet collision
goes like $\log(k_T)/k_T^4$ while that from the quark acceleration goes like $1/k_T^2$.
Different variation in different ranges of $k_T$ is discussed, for example, in \cite{dumitru}.
\item The rapidity distribution of the quark-sheet collision includes 
a target fragmentation region for $y<\log(k_T/m)$, $m=$ IR regulator, within which 
the distribution shows a linear increase, Fig.~\ref{N2(y)}, beyond which there is a plateau extending
arbitrarily. The distribution following from quark acceleration has a natural large $y$ cut-off given by the
momentum of the accelerated quark. 
\end{itemize}

At first sight, the different $k_T$ dependences of these two processes seems alarming.  It should not be so.  
The $k_T$ dependence of $1/k_T^2$ is only valid for the direct charged particle emission and in the 
range where $k_T  \ll Q_{\rm sat}^{\rm proj}$, that is when the typical transverse momentum of the emitted gluon 
is small compared to the typical momentum kick the charged particle gets from scattering from the nucleus.  
At higher momentum, we simply must modify the computation to take into account the charged particle recoil.  
In this region the production cross section will fall like $1/k_T^4$, although our method of computation  
presented here will fail in this region.

For the Gunion-Bertsch contribution, we explicitly worked in the region where $k_T \gg Q_{\rm sat}^{\rm proj}$.    
In this region the non-linearities of the
projectile color field are unimportant. Our approximations are valid in this region since no large recoil of the charged 
particle is required, and this region provides a useful check of our computations.

The interesting region of computation is when the $Q_{\rm sat}^{\rm targ}  \ll k_T  \ll Q_{\rm sat}^{\rm proj}$.  This is where most of the 
particle production takes place.  In this region, the Gunion-Bertsch computation is not sufficient, and the full non-linearity
of the projectile color field must be properly taken into account, i.e., a more accurate evaluation of the expectation value in
Eq.~(\ref{finres}) is needed.

\begin{figure}[!t]
\begin{center}
\includegraphics[width=0.5\textwidth]{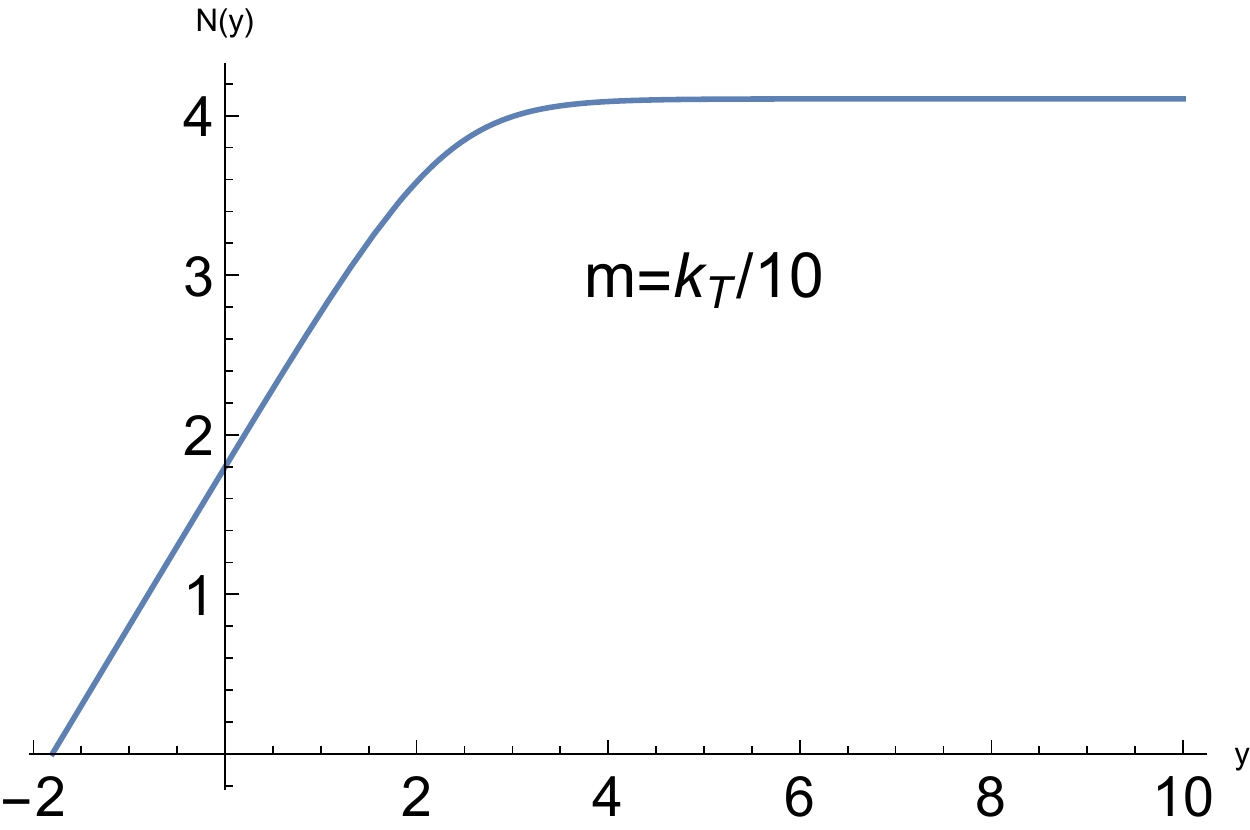}
\caption{\small  Rapidity distribution (bracketed factor in Eq.~\protect(\ref{sheety}) is plotted) at fixed $k_T$ of gluons 
emitted in a collision of an initially static quark with 
nucleus. The IR regulator mass is $k_T/10$ and the curve shows a linear increase as a function of rapidity in
the fragmentation region. }
\label{N2(y)}
\end{center}
\end{figure}

 \section{Conclusions}\label{concl}
We have in this article computed gluon production in a collision of an ultrarelativistic nucleus and a static
quark. The result consists of two parts, an ED-like inhomogeneous contribution from quark acceleration (Eqs.~(\ref{mult1},\ref{Jeff}) and
a homogeneous contribution from the interaction between the nuclear sheet and the quark (Eq.~(\ref{finres})). This interaction
term corresponded to a homogeneous solution of the radiation equation due to a special gauge choice (\ref{U}) in which
the gauge field vanished after the passage of the sheet.

This is only the first step towards the final goal, making predictions for target fragmentation dynamics of ultrarelativistic
nuclear collisions. The next step involves performing the transverse momentum convolution and the color ensemble
averaging in Eq.~(\ref{finres}). Here they were carried out only for dilute systems. Next these gluonic results should be combined with
those for quarks in \cite{McLerran:2018avb} to give initial values for energy-momentum and baryon number. This is analogous
with the very early work in \cite{km}. Finally, one should numerically go through hydrodynamic evolution in analogy 
with \cite{Kajantie:1982zb} and prepare predictions for experiments - which hopefully some day will well extend to the
fragmentation region.

\begin{acknowledgments}
K. Kajantie and R. Paatelainen thank Tuomas Lappi and Mark Mace for discussions. L. McLerran acknowledges a useful discussion
with Bjoern Schenke and Chun Shen that rekindled his interest in this problem.  All of the authors thank Aleksi Vuorinen and Aleksi Kurkela 
for organizing the Workshop on Hot and Dense QCD in Saariselka,  Finland where this collaboration was initiated.
R. Paatelainen is supported by the European Research Council, grant no. 725369 and L.  McLerran was supported by 
the U.S. DOE under Grant No. DE-FG02- 00ER41132.
\end{acknowledgments}

\end{document}